\documentclass[3p,times,twocolumn]{elsarticle}

\usepackage{ecrc}

\usepackage{natbib}
\usepackage{graphicx}
\usepackage{epstopdf}

\volume{00}

\firstpage{1}

\journalname{Nuclear and Particle Physics Proceedings}

\runauth{V. Zekovi\'c \& B. Arbutina}

\jid{nppp}

\jnltitlelogo{Nuclear and Particle Physics Proceedings}

\usepackage{amssymb}

\usepackage{amsthm}

\usepackage[figuresright]{rotating}

\usepackage[percent]{overpic}

\begin{document}

\begin{frontmatter}



\dochead{}

\title{Resonant Particle Backscattering of a Shock Wave}


\author[label1]{Vladimir Zekovi\'c\corref{cor1}}
\author[label1]{Bojan Arbutina\corref{cor2}}

\address[label1]{Department of Astronomy, Faculty of Mathematics, 		University of Belgrade\\
		Studentski trg 16, 11000 Belgrade, Serbia}
		
\cortext[cor1]{vlada@matf.bg.ac.rs}
\cortext[cor2]{arbo@matf.bg.ac.rs}

\begin{abstract}
A collisionless shock wave is treated as a streaming plasma instability in the interstellar medium (ISM). We assume that in a steady state, this instability propagates through the ISM as a self-driven plasma instability, whereby the parameters of the instability are determined such that it causes density and velocity jumps as well as isotropization of the particle velocities, which then must be in accordance with MHD theory. The process of resonant interaction and particle scattering off such instability and their backscattering to an upstream region is simulated. We find that some ISM particles bounce off a shock and thus become suprathermal and eligible to enter the process of acceleration to cosmic ray energies by the mechanism of diffusive shock acceleration.
\end{abstract}


\end{frontmatter}


\section{Introduction}
\label{intro}

It is now known from particle-in-cell (PIC) simulations~\citep{SDA} that the \emph{non-resonant} modes of Bell's instability~\citep{NR}, which are seeded by return current of CRs, pre-accelerate and return ions through the process of shock drift acceleration (SDA). Acceleration is governed by the electric field component associated with these modes, while magnetic field has negligible influence due to wavelengths typically much shorter then ion gyroradii. \emph{Resonant} modes can grow faster than \emph{non-resonant} modes depending on the Alfv\'enic Mach number of a shock~\citep{RvsNR}. In either case, such instabilities are bound to region ahead of a shock.

Here, we consider a different type of instabilities which arise when one plasma interpenetrates another. In their nature, such instabilities are Alfv\'enic, and by the way of excitation, they are streaming or flow-driven instabilities~\citep{FDM} that draw the energy for their growth from the macroscopic kinetic energy pool of two interpenetrating plasmas. Depending on the plasma density, their motion is more or less coupled to the flow, meaning that they should be bound to the shock itself and to the region behind it. As seen in a shock frame, a mode that is in resonance with the stream of an incoming plasma is growing faster then non-resonant modes, and is also favoured by the micro-physics of wave-particle interactions. Beacuse of its resonant nature, instability scatters particles strongly, and by that it produces density jumps and isotropization of particle velocities. It also backscatters particles into an upstream region, thus contributing to the return current.

We simulate these processes by a pure electromagnetic N-body code, to study the dynamics of particles whose longitudinal Larmor radii are in the band of resonant wavelengths, on the micro-scale of already formed instability. From that, we want to point out the main properties of such resonant interaction and also to make assumptions on how such a dynamics manifests itself on scale of a shock.

We organize this paper as follows. In Section 2, we first describe the properties of streaming plasma instabilities and derive condition for an \emph{in-stream} resonance. We present results of electromagnetic N-body simulations in Section 3. In Section 4, we present an overall discussion, and also present preliminary results of long-term two-dimensional PIC simulations.


\section{Streaming Alfv\'en instability}
\label{resonant}

We consider two interpenetrating ion-electron plasmas with same ion species, where a flowing plasma streams with speed $\varv_0$ along the parallel lines of a background magnetic field $\vec{B_0}$ embedded in another, static plasma. As derived in \citep{FDM}, momentum equations in the low-frequency limit ($\omega \ll \Omega_i = e B_0 / m_i$) are
\begin{equation}
\rho_s \frac{\partial \vec{u_1}}{\partial t} = \vec{j_{s1}} \times \vec{B_0},
\end{equation}
\begin{equation}
\rho_f \left( \frac{\partial}{\partial t} + \varv_0 \cdot \nabla \right)\ \vec{\varv_1} = \vec{j_{f1}} \times \vec{B_0},
\end{equation}
\noindent where $\vec{u_1}$ and $\vec{\varv_1}$ are the perturbed velocities of the static and flowing plasmas respectively. Densities $\rho_x = (m_i + m_e) n_x$, perturbed currents $\vec{j_{x1}}$ and all other symbols in equations, are denoted with $f$ for flowing and $s$ for static plasma. Because both plasmas are collisionless, the coupling between them is achieved through the perturbed magnetic field, as stated by the Amp\`ere law

\begin{equation}
\nabla \times \vec{B_1} = \mu_0\ (\vec{j_{f1}} + \vec{j_{s1}}).
\end{equation}
After applying conditions $\vec{E_1} + \vec{u_1} \times \vec{B_0} = 0$ for static and $\vec{E_1} + \vec{\varv_1} \times \vec{B_0} + \vec{\varv_0} \times \vec{B_1} = 0$ for moving plasma, in (1) and (2), currents are derived and then used in (3). Solution to (3) is then dispersion relation, which shows that Alfv\'en wave can couple with the flow-driven $\omega = k \varv_0$ fluid mode. In frame of a steady plasma solution is

\begin{equation}
\omega^{2} \left( 1 + \frac{\varv_{Af}^2}{\varv_{As}^2} \right) - 2 k \varv_{0} \omega + k^{2} (\varv_{0}^2 - \varv_{Af}^2) = 0,
\end{equation}

\noindent where $\varv_A$ is an Alfv\'en speed . We also define $\eta=n_f/n_s$ as the number density ratio between the two plasmas. The solutions that emerge from this relation are

\begin{equation}
\frac{\omega}{k} = \frac{1}{1+{1 /\eta}} \cdot \left\{ \varv_0 \pm \varv_{As}\sqrt{{1 \over \eta} \left( 1+\frac{1}{\eta}-\frac{\varv_{0}^2}{\varv_{As}^2} \right)} \right\},
\end{equation}

\noindent one with the plus sign which propagates in the direction of a flow and one with the minus sign which propagates backward, P and N modes respectively. For a large enough flow speed, these modes may merge and form an instability on the condition that

\begin{equation}
\varv_{0}^2 > \varv_{As}^2 \left( 1 + \frac{1}{\eta} \right) = \varv_{As}^2 + \varv_{Af}^2.
\end{equation}

\noindent The solution thus becomes complex $\omega = \Omega_r \pm i \Gamma$ with the positive imaginary part as the growth rate

\begin{equation}
\Gamma = \frac{k}{1+{1 /\eta}} \sqrt{\frac{1}{\eta} \left| \varv_{As}^2  + \varv_{Af}^2 - \varv_{0}^2 \right|}\ .
\end{equation}

Modes with shorter wavelengths grow faster and are favoured by these equations. However, circular polarization has to be considered for wavelengths shorter than or equal to the resonant wavelength $\lambda \lesssim r_g = \gamma m_i \varv_0 / q B_0$. By looking at the microphysics of these processes, we can see that particles do not 'feel' modes with wavelengths shorter than the resonant wavelength. Particles interact very weakly with them, and only wiggle a little bit around their equilibrium position while streaming along the background field (Figures 1 and 2). Long-wavelength modes confine particles, so their guiding centers follow the perturbed field lines. As a growth rate increases linearly with wavenumber, \emph{resonant} modes grow faster than long-wavelength modes. Because of their resonance with longitudinal particle gyro-radii, these modes should quickly disrupt the stream, and by that, also damp short-wavelength \emph{non-resonant} modes.

\section{Resonant interaction with particles}
\label{interaction}

We now apply results from \citep{FDM} and our assumptions of an \emph{in-stream} resonance, to case of a parallel shock. We assume that the formation of an instability should start with an explosion. In case of a supernova explosion, the ejecta expanding with velocity $\varv_0$, i.e., the flowing plasma, is much denser than static, surrounding interstellar plasma, so $\eta = n_f / n_s \gg 1$. If $\varv_0 > \varv_{As}$, the dispersion relation (5) then becomes
\begin{equation}
\frac{\omega}{k} \approx \varv_0 + i \sqrt{{1 \over \eta} \left( \varv_{0}^2 - \varv_{As}^2 \right)}.
\end{equation}
\noindent In this case, an Alfv\'en wave is strongly coupled to the flow and forms an instability that moves with the ejecta.

We run our simulations in a frame of ejecta or downstream frame, where the instability is static. To optimize computing power, the electric field is neglected and only protons are considered. The velocity distribution is assumed to be Maxwellian for a temperature $T_0 = 10^4 \rm K$, shifted by the bulk speed of the flow, $\varv_0$.

Here, we investigate the parameter space, by simultaneously varying the amplitude (Figure 1) and wavelength (Figure 2) of an instability, to see what kind of dynamics such a \emph{resonant} instability imposes on streaming particles.

The graphs in Figure 1 show separate cases of purely circular and purely linear polarizations. We see that the interaction strength changes non-linearly with instability amplitude, in a way that scattering is triggered when the transverse field component reaches an amplitude $B_{\perp} \approx B_0$. Even for such a small amplitude, incoming particles are scattered in pitch angle and their mean longitudinal velocity drops significantly, thus contributing to the formation of a shock region where density jumps, as well as isotropization of particle velocities, occur. For amplitudes greater than $B_0$, incoming particles completely bounce off an instability, draining the energy from it, and possibly constraining its amplitude to grow not larger than $B_0$.

In Figure 2, we show phase space diagrams of a flow interacting with instabilities of different wavelengths. It can be seen that the strongest interaction is within a narrow band of resonant wavelengths, $\lambda_{res} \sim \pi r_g$.

\begin{figure}[!h]
	\vskip -5mm
	\centering
	\begin{overpic}[width=\linewidth]{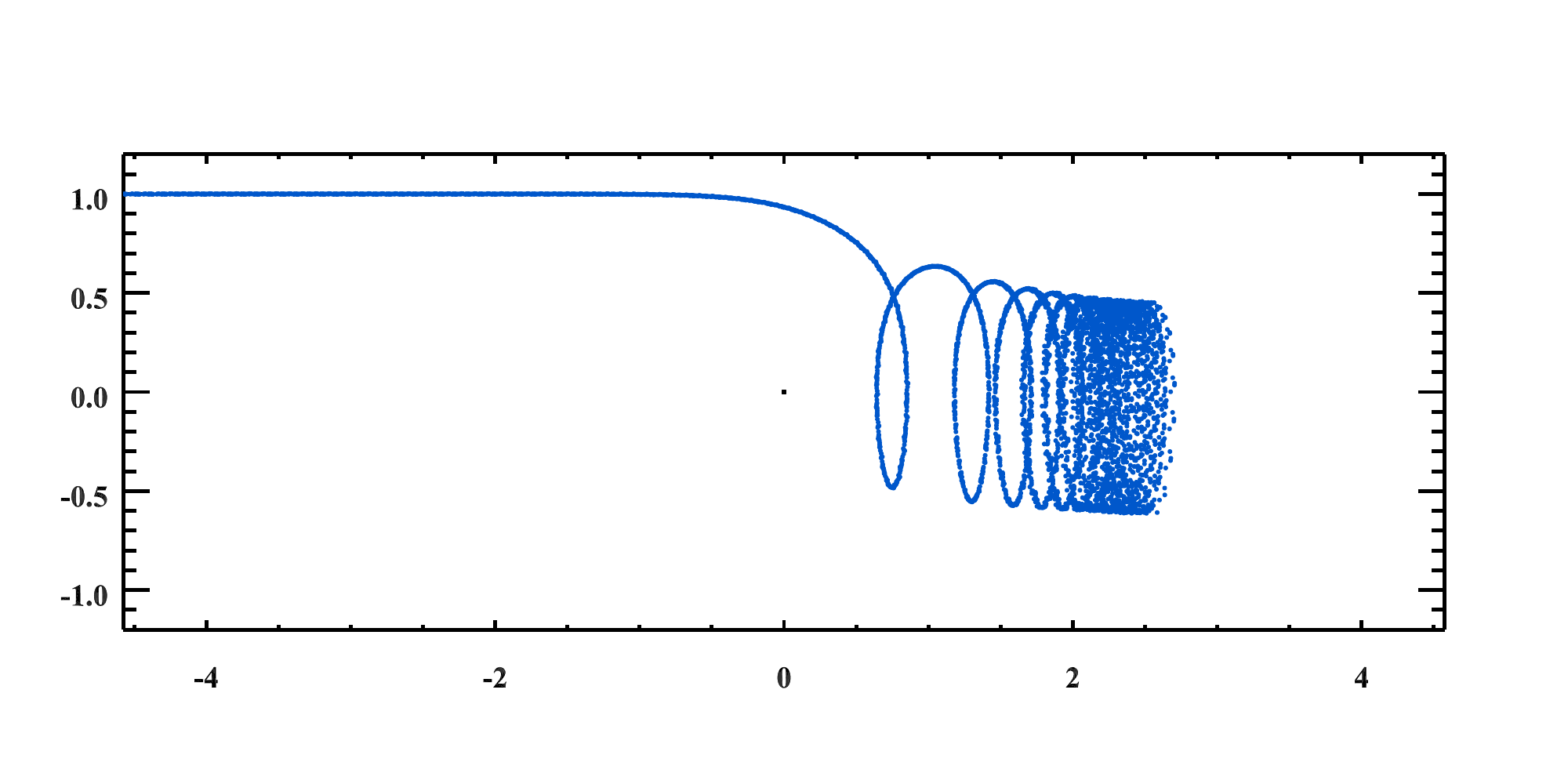}
		\put(0,0){\includegraphics[width=\linewidth]{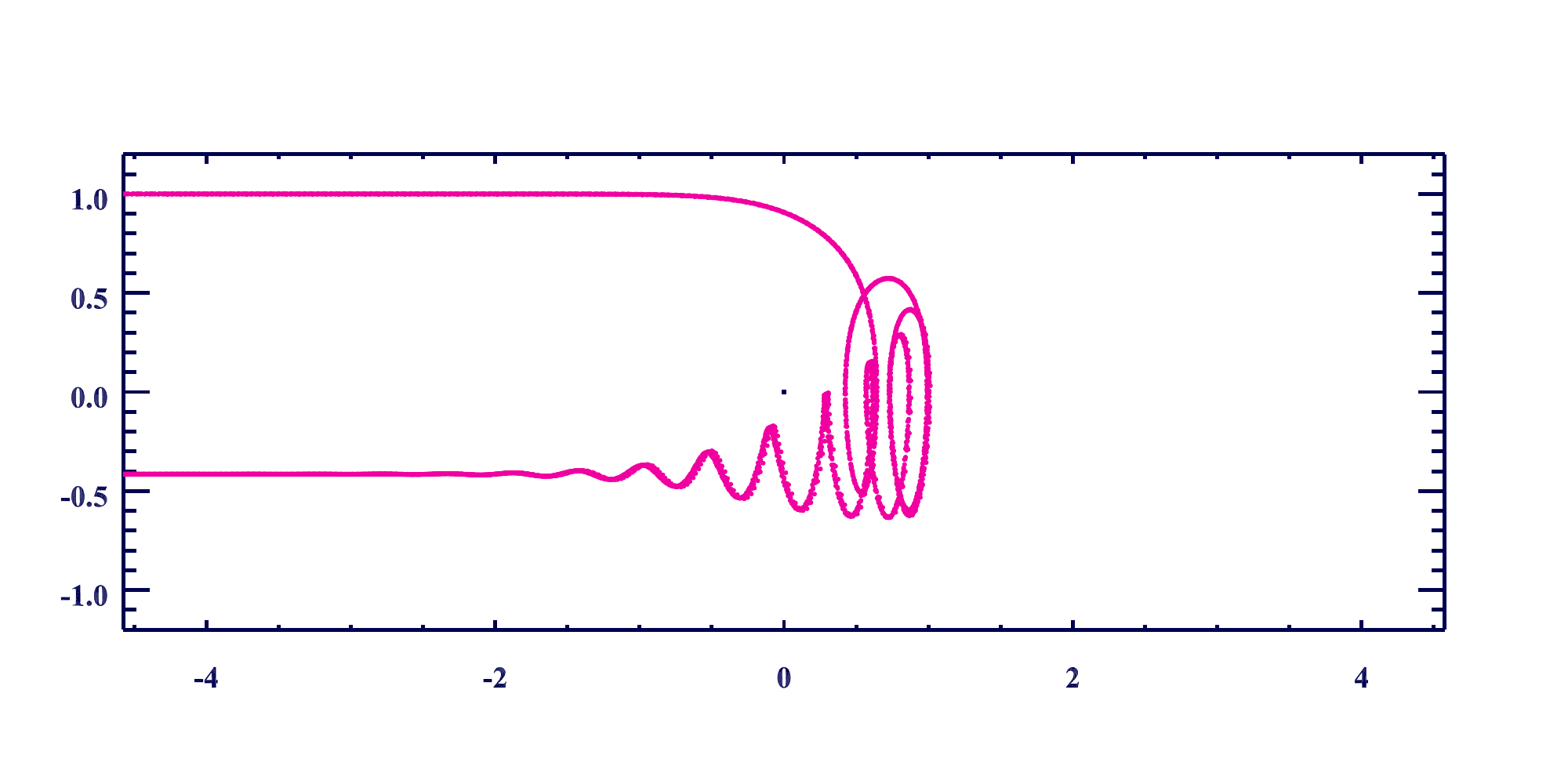}}
		\put(0,0){\includegraphics[width=\linewidth]{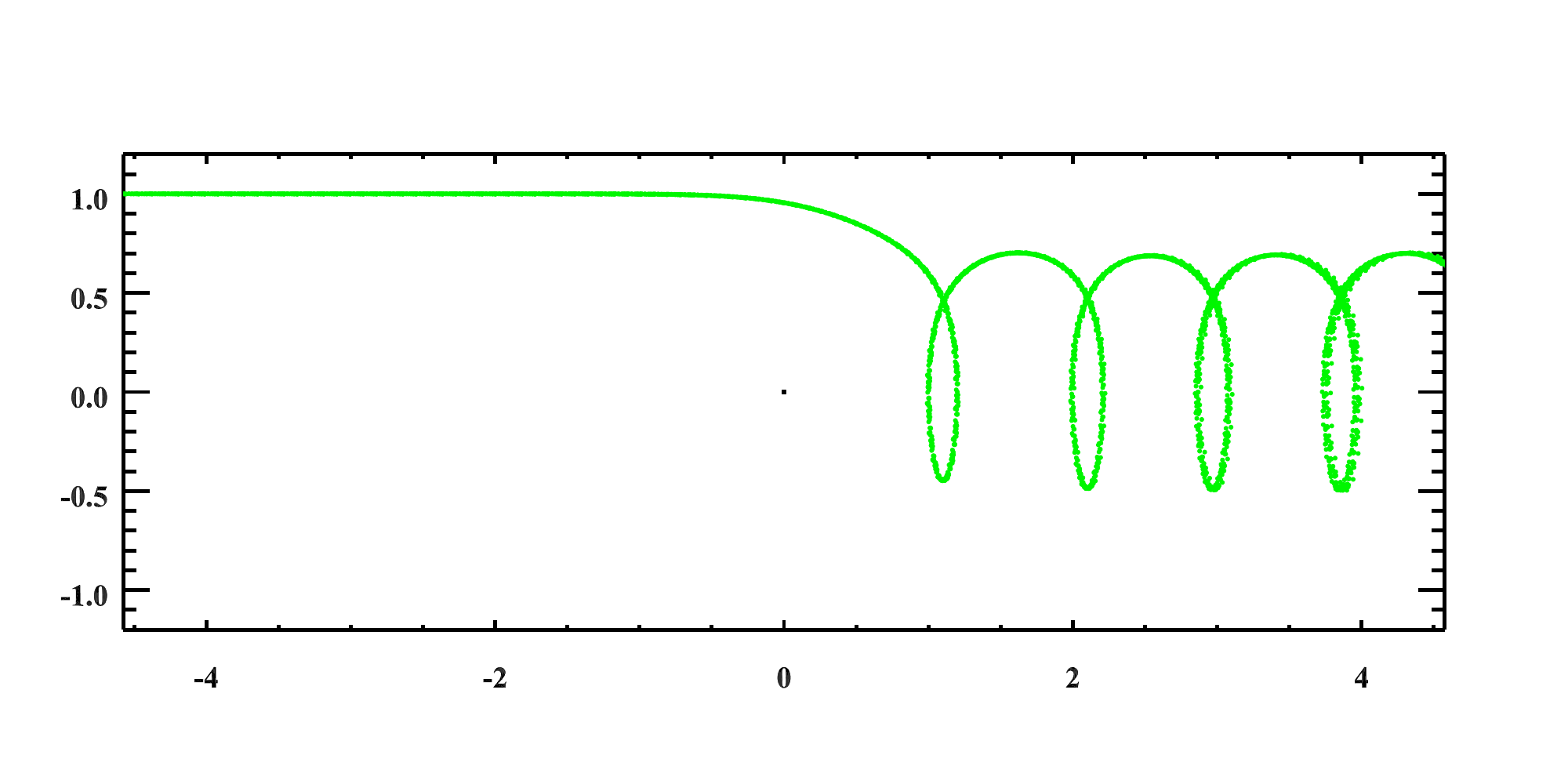}}
	\end{overpic}
	\vskip -10mm
	\begin{overpic}[width=\linewidth]{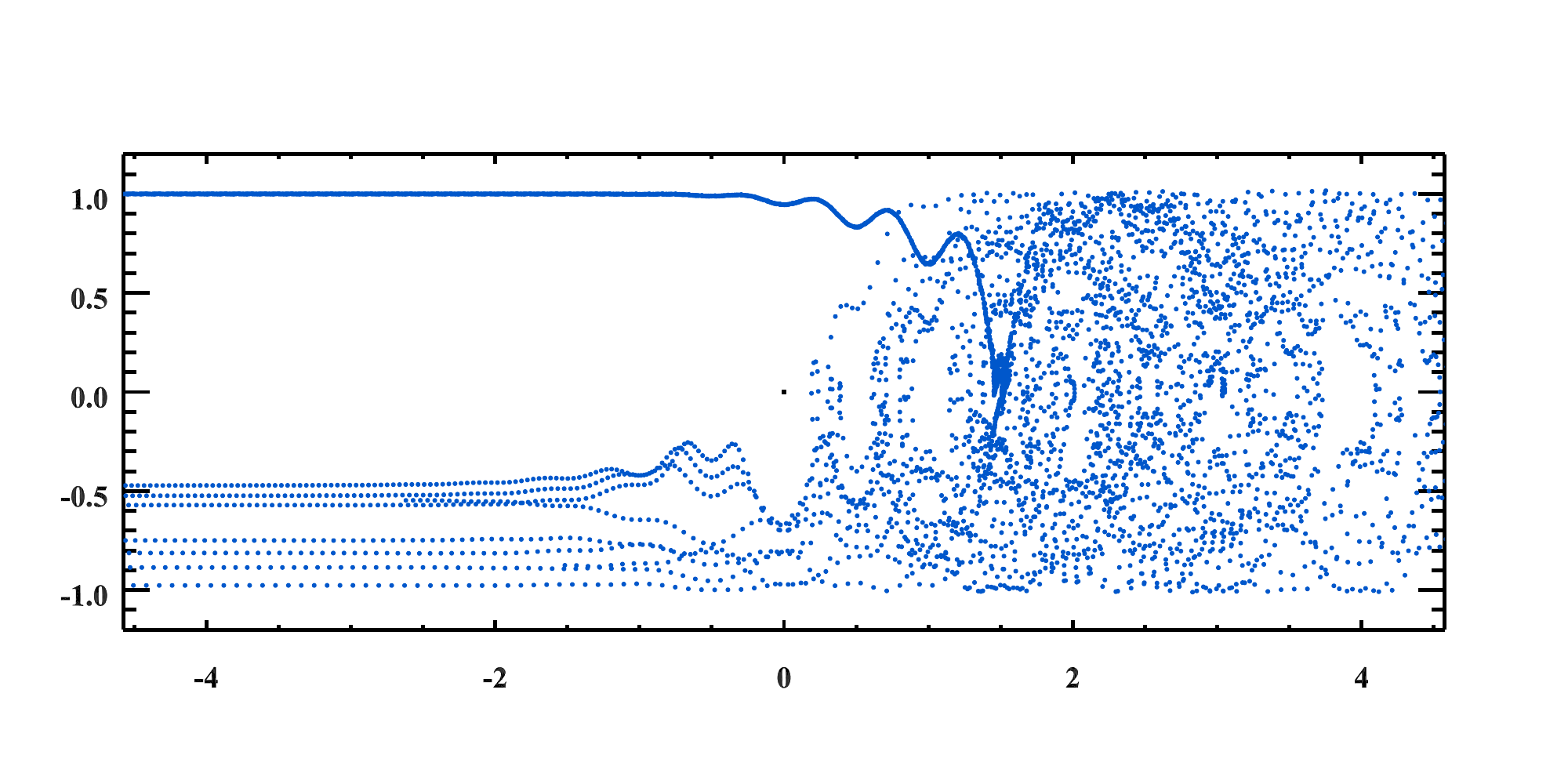}
		\put(0,0){\includegraphics[width=\linewidth]{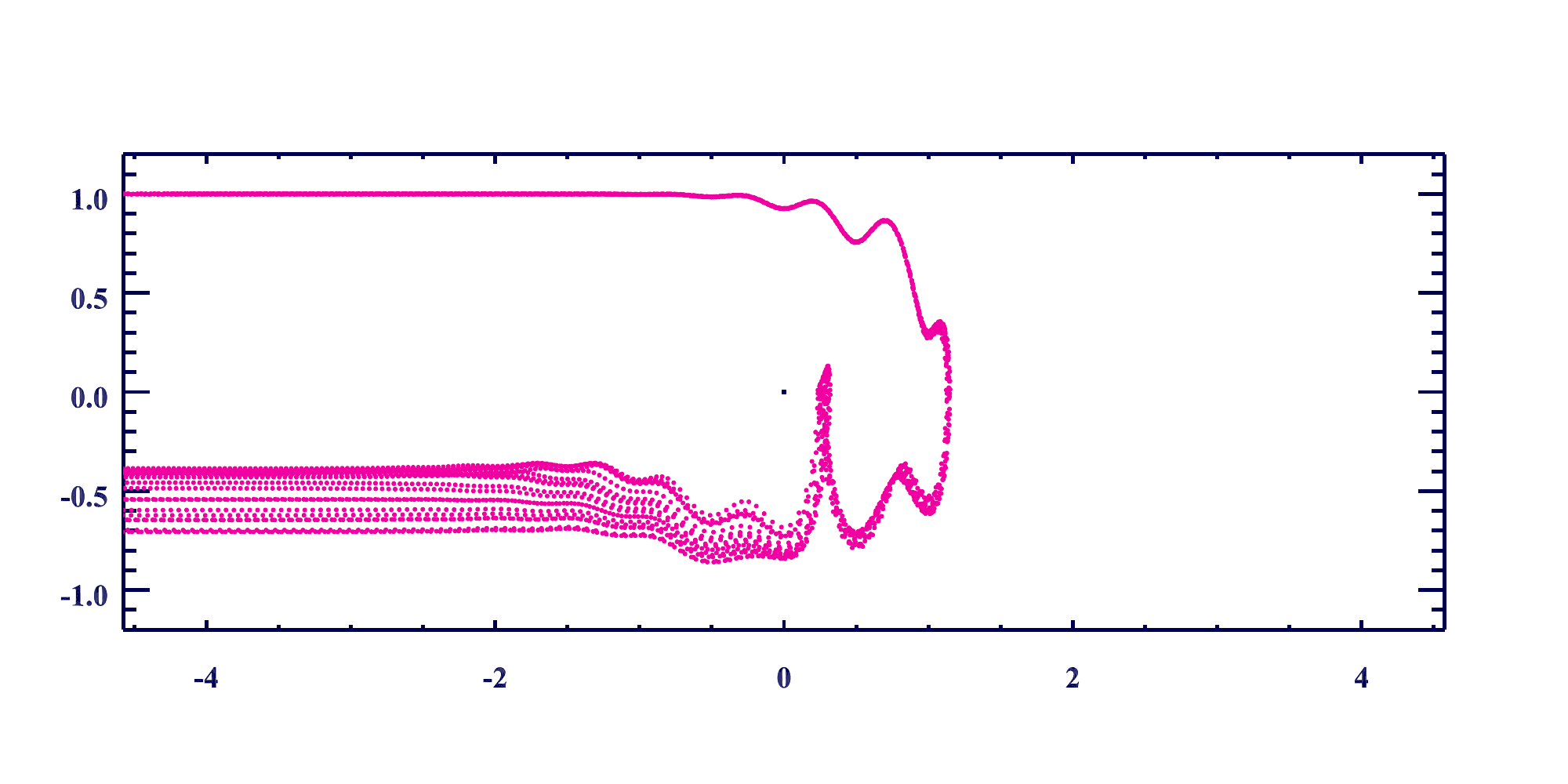}}
		\put(0,0){\includegraphics[width=\linewidth]{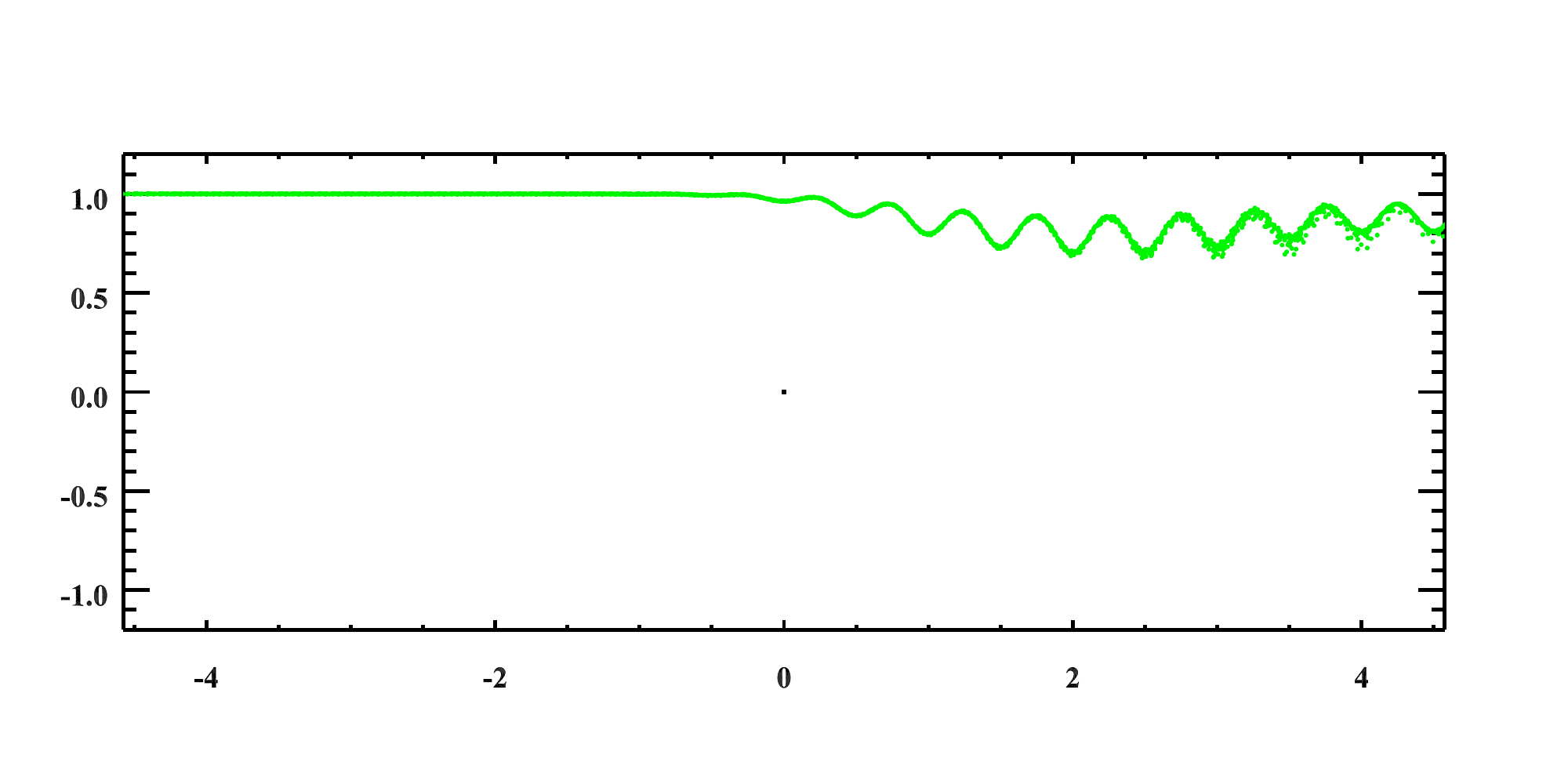}}
	\end{overpic}
	\vskip -3mm	
	\caption{Phase space diagrams of circularly (top) and linearly (bottom) polarized instabilities interacting with the flow of particles. On the horizontal axes the X-coordinates are given in units of gyro-wavelength ($\lambda_{g} = \pi r_g$); and on the vertical axes, the longitudinal velocities are normalized to $\varv_0$. Colors denote different instability amplitudes: $0.83\ B_0$ (green), $B_0$ (blue), and $1.17\ B_0$ (purple). Particles stream left to right.}
\end{figure}

We also observe that both polarizations separately, cause the slowing down of an incoming plasma. However, concerning circular mode of polarization, there is a significant drop of collective longitudinal motion, but with no particles backscattered to an upstream region. On the other hand, the linear mode generally does not stop the plasma, but rather it causes the plasma to slowly drift. But in this case the upstream scatter of particles is very strong.

\begin{figure}[!h]
\centering
	\vskip -5mm
	\begin{overpic}[width=\linewidth]{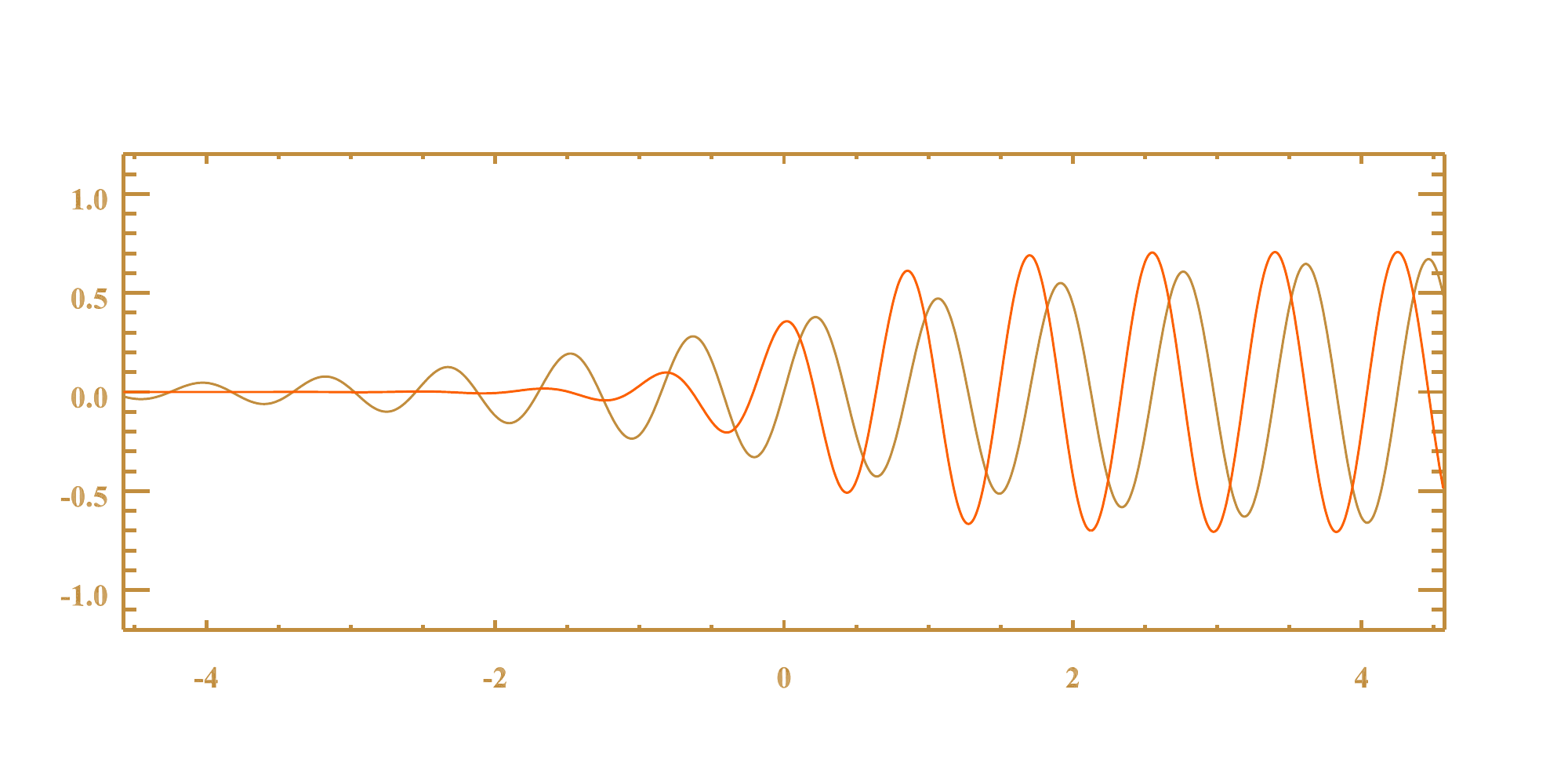}
		\put(0,0){\includegraphics[width=\linewidth]{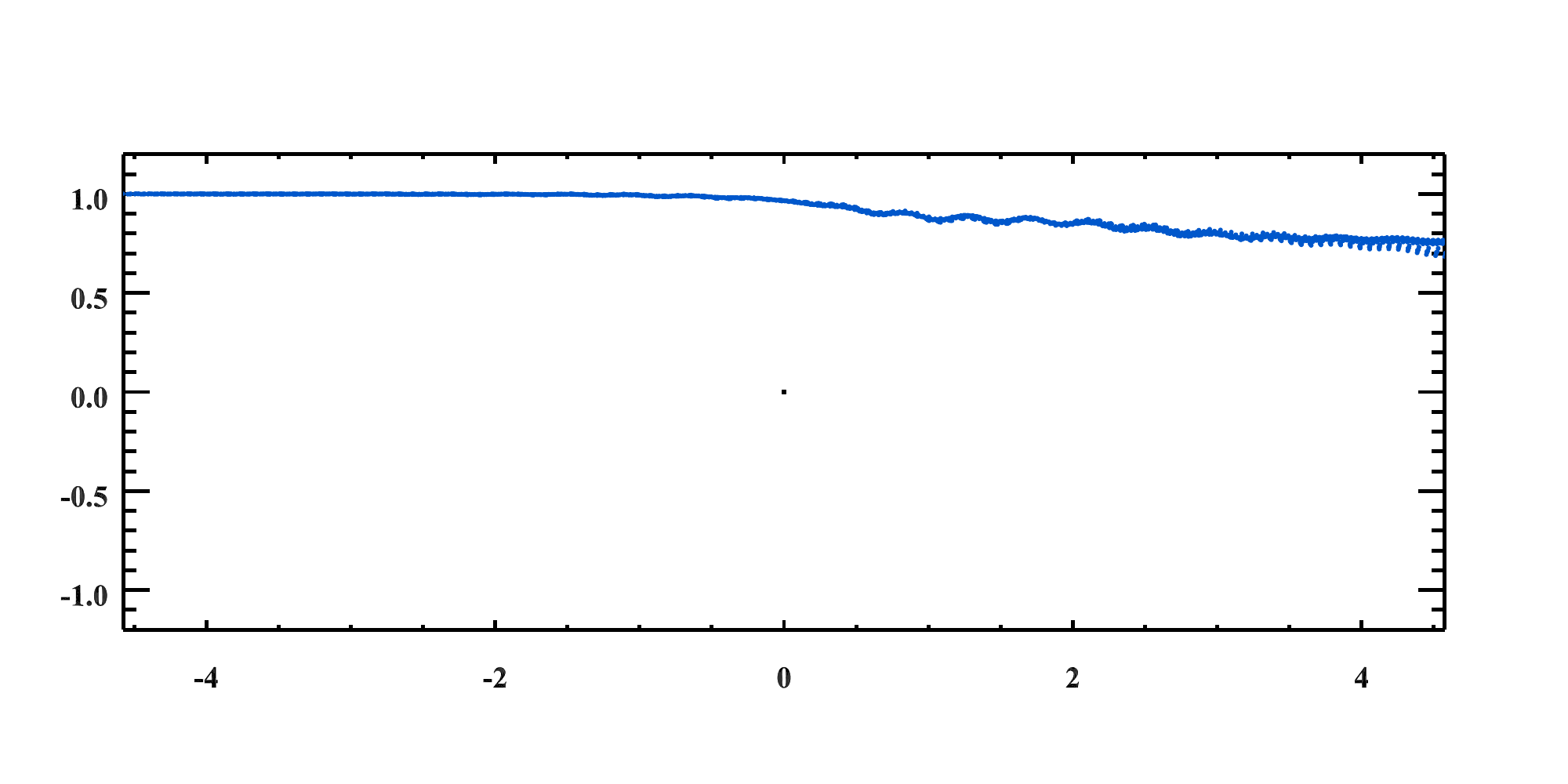}}
	\end{overpic}
	\vskip -11mm
	\begin{overpic}[width=\linewidth]{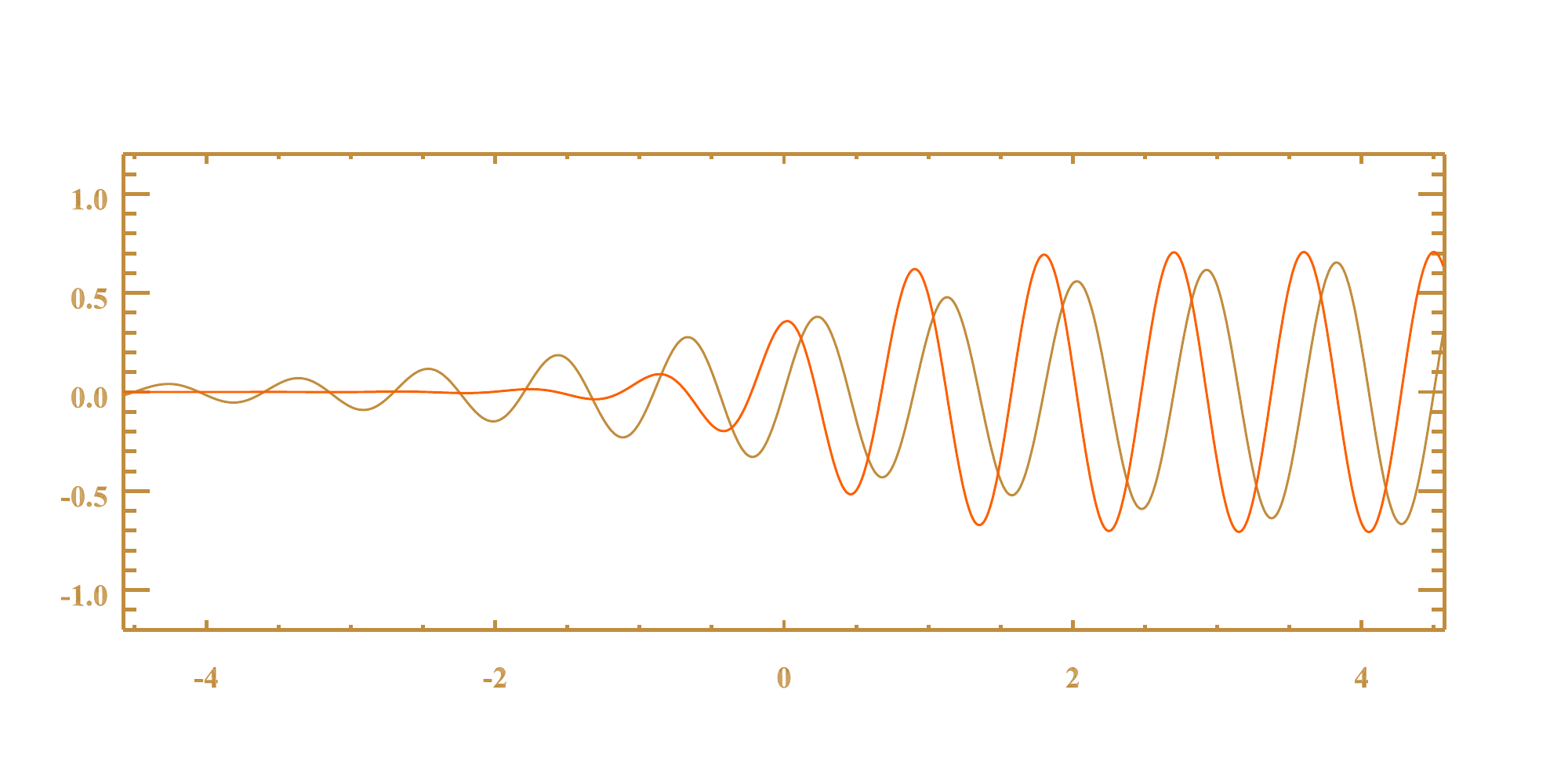}
		\put(0,0){\includegraphics[width=\linewidth]{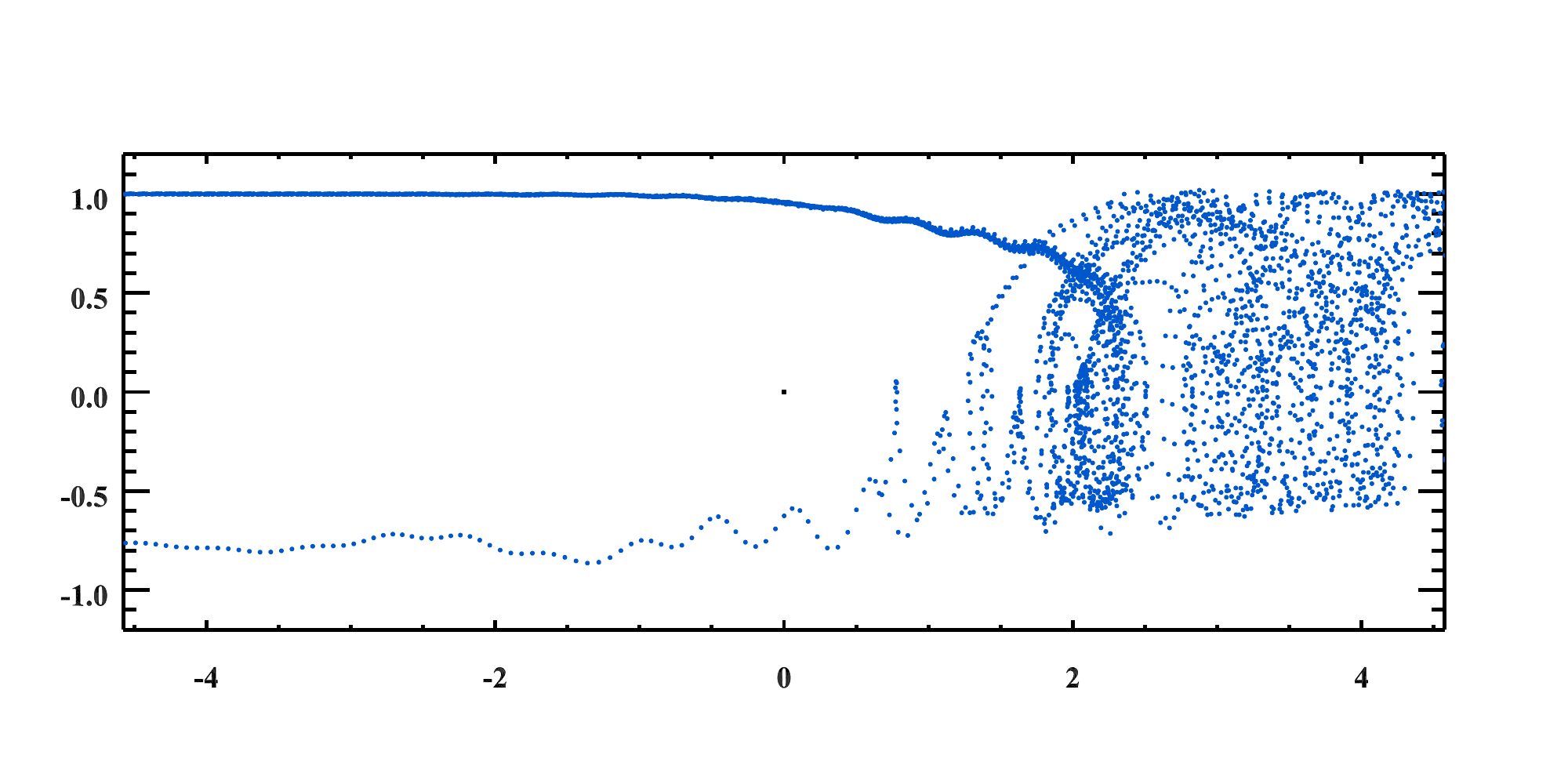}}
	\end{overpic}
	\vskip -11mm
	\begin{overpic}[width=\linewidth]{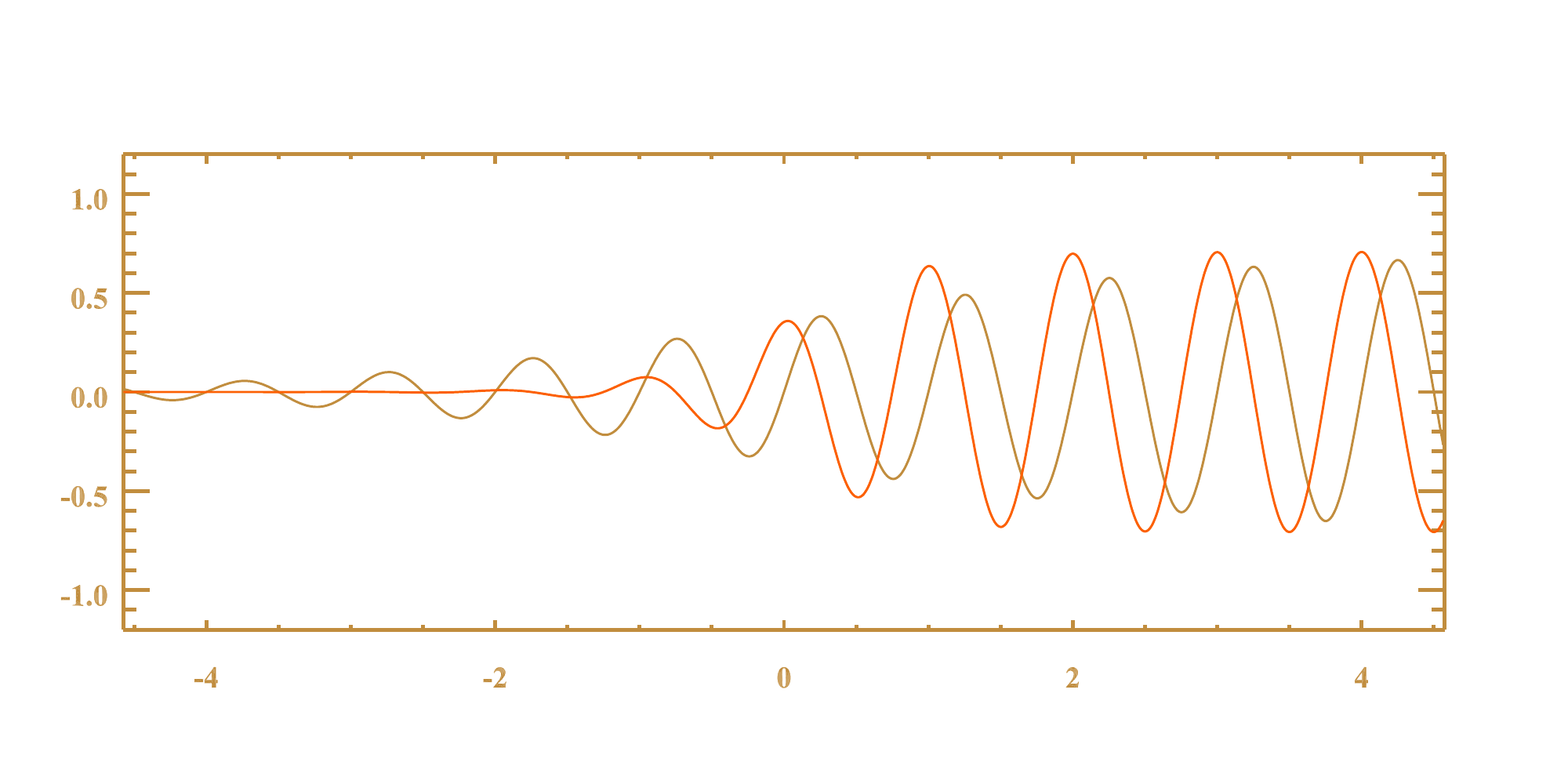}
		\put(0,0){\includegraphics[width=\linewidth]{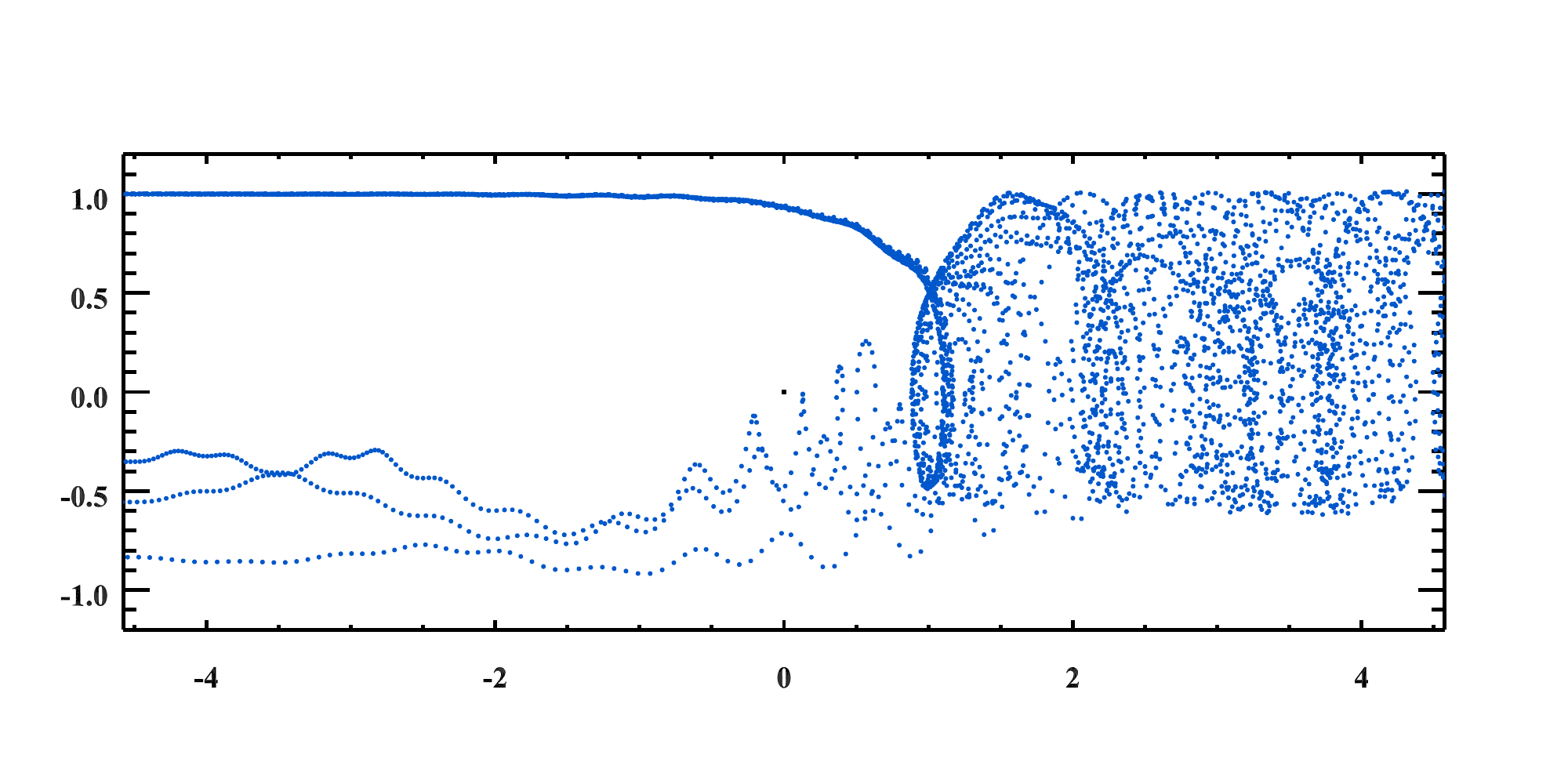}}
	\end{overpic}
	\vskip -11mm
	\begin{overpic}[width=\linewidth]{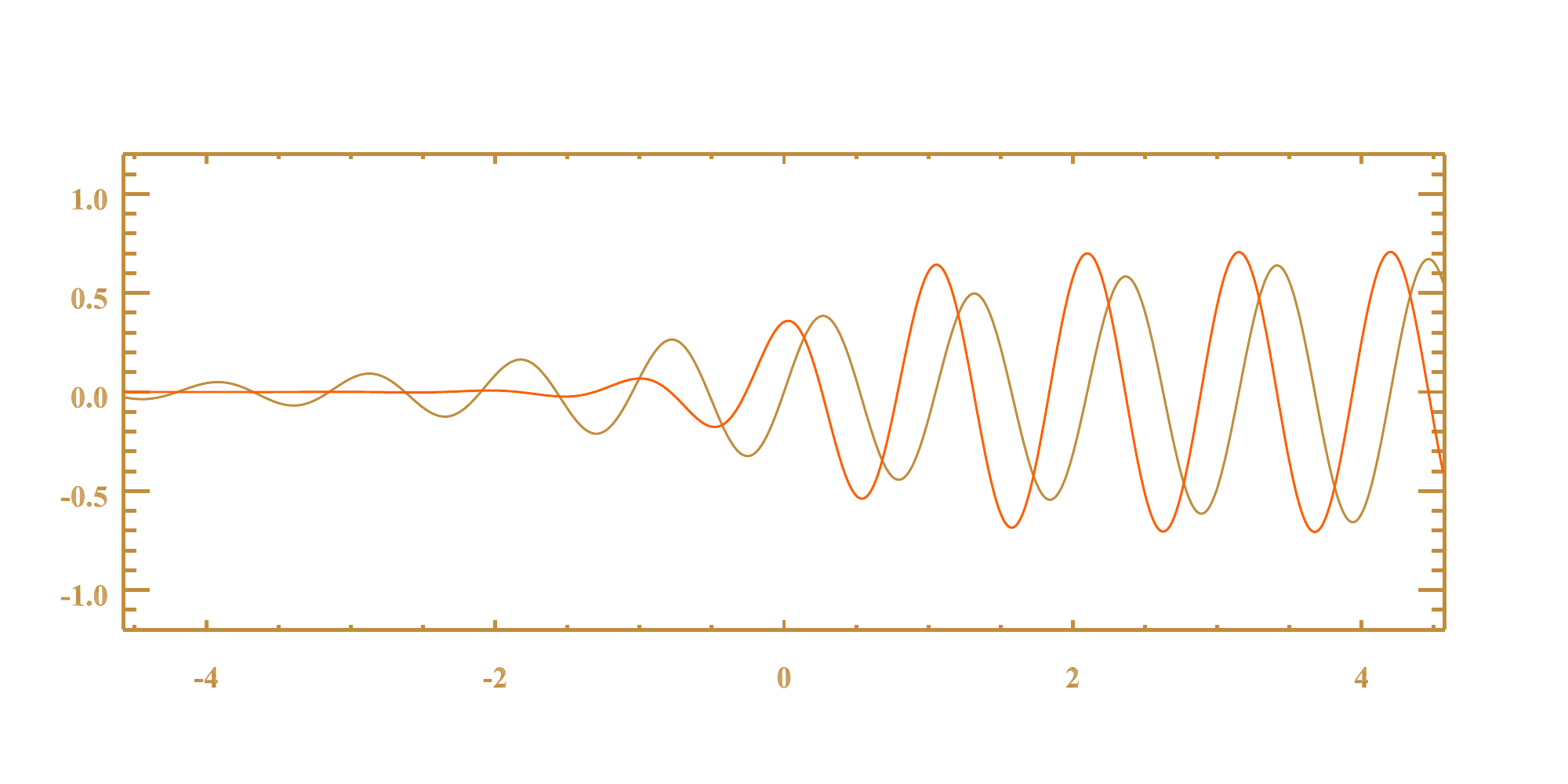}
		\put(0,0){\includegraphics[width=\linewidth]{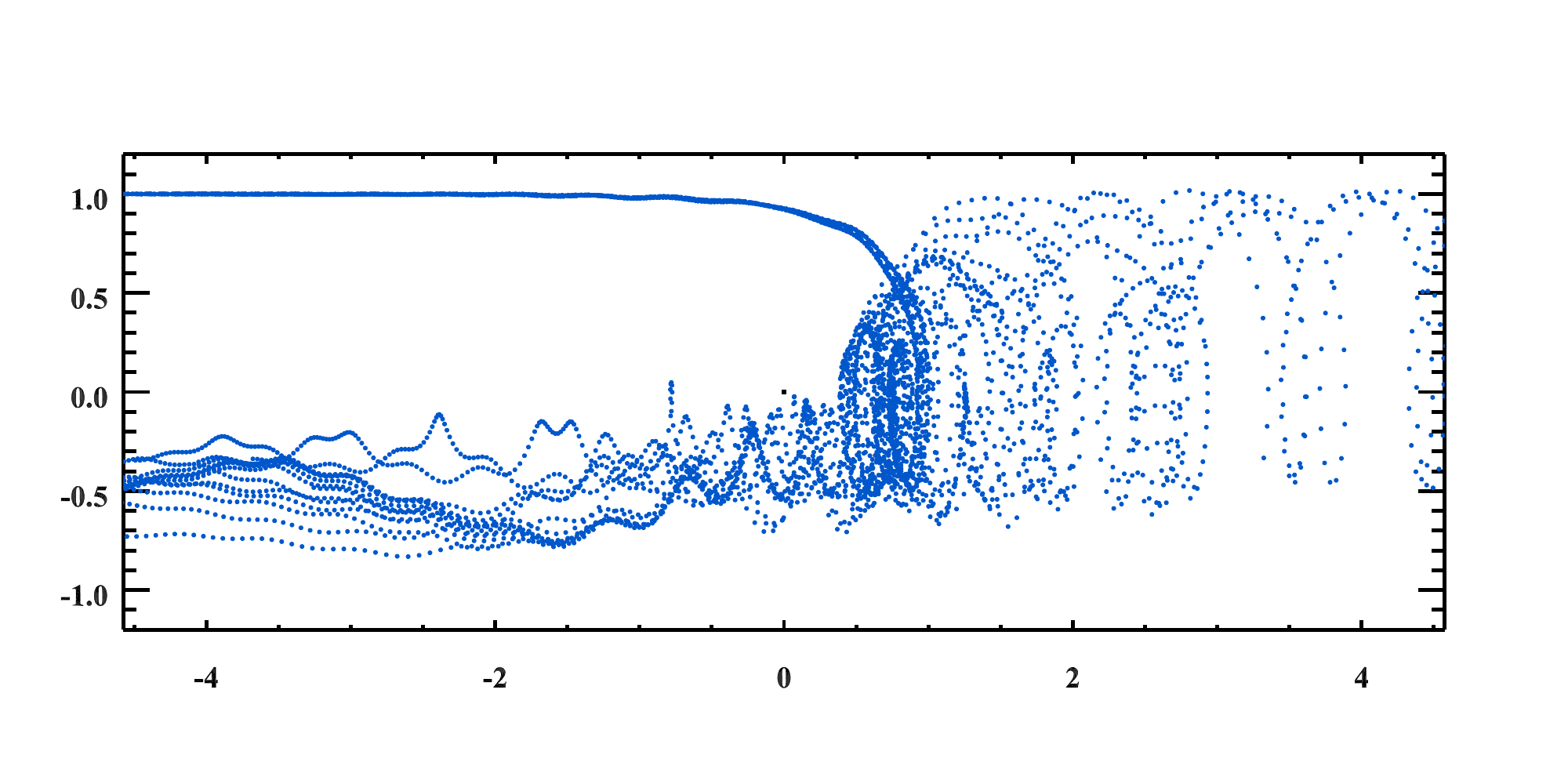}}
	\end{overpic}
	\vskip -11mm
	\begin{overpic}[width=\linewidth]{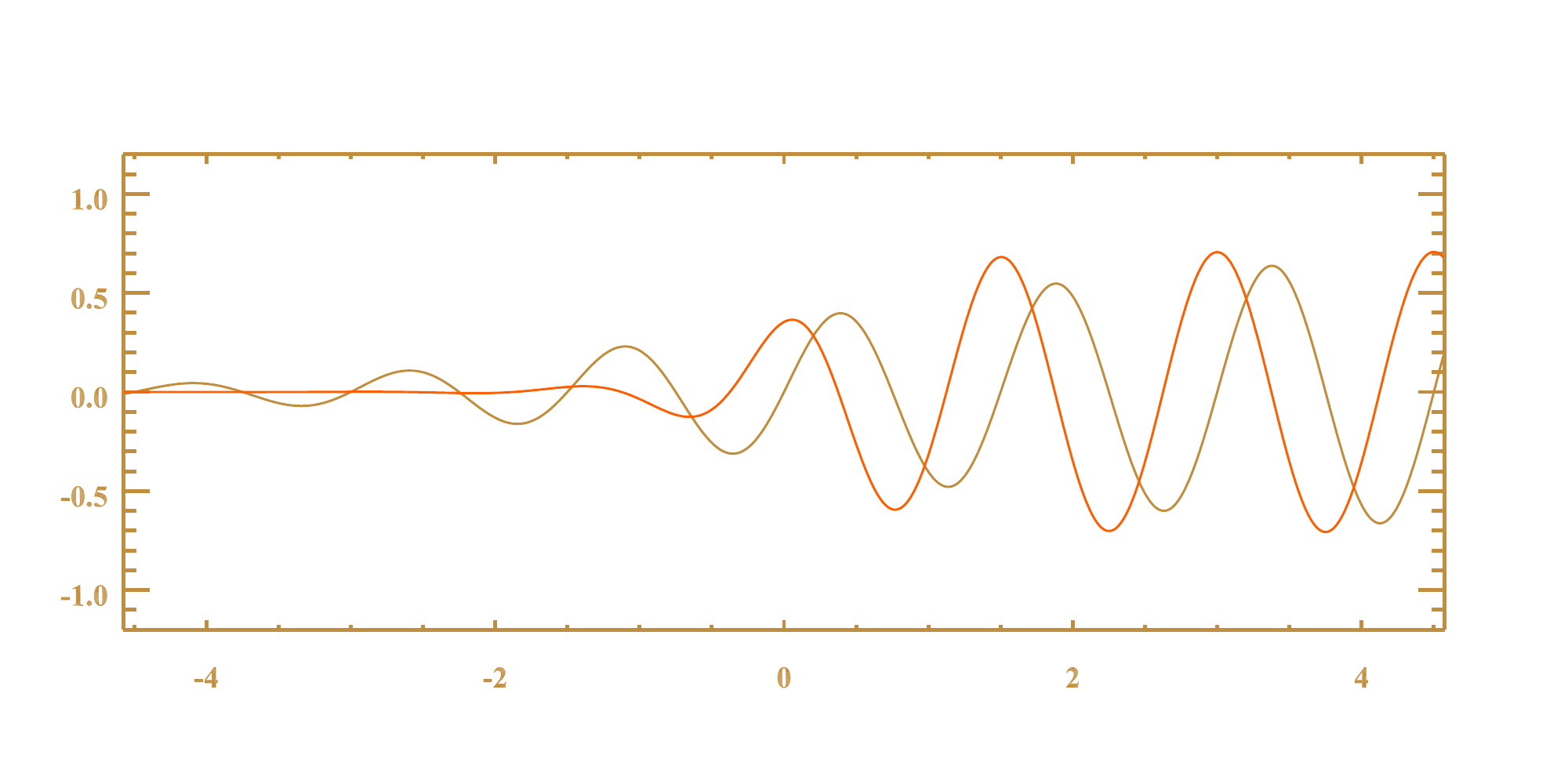}
		\put(0,0){\includegraphics[width=\linewidth]{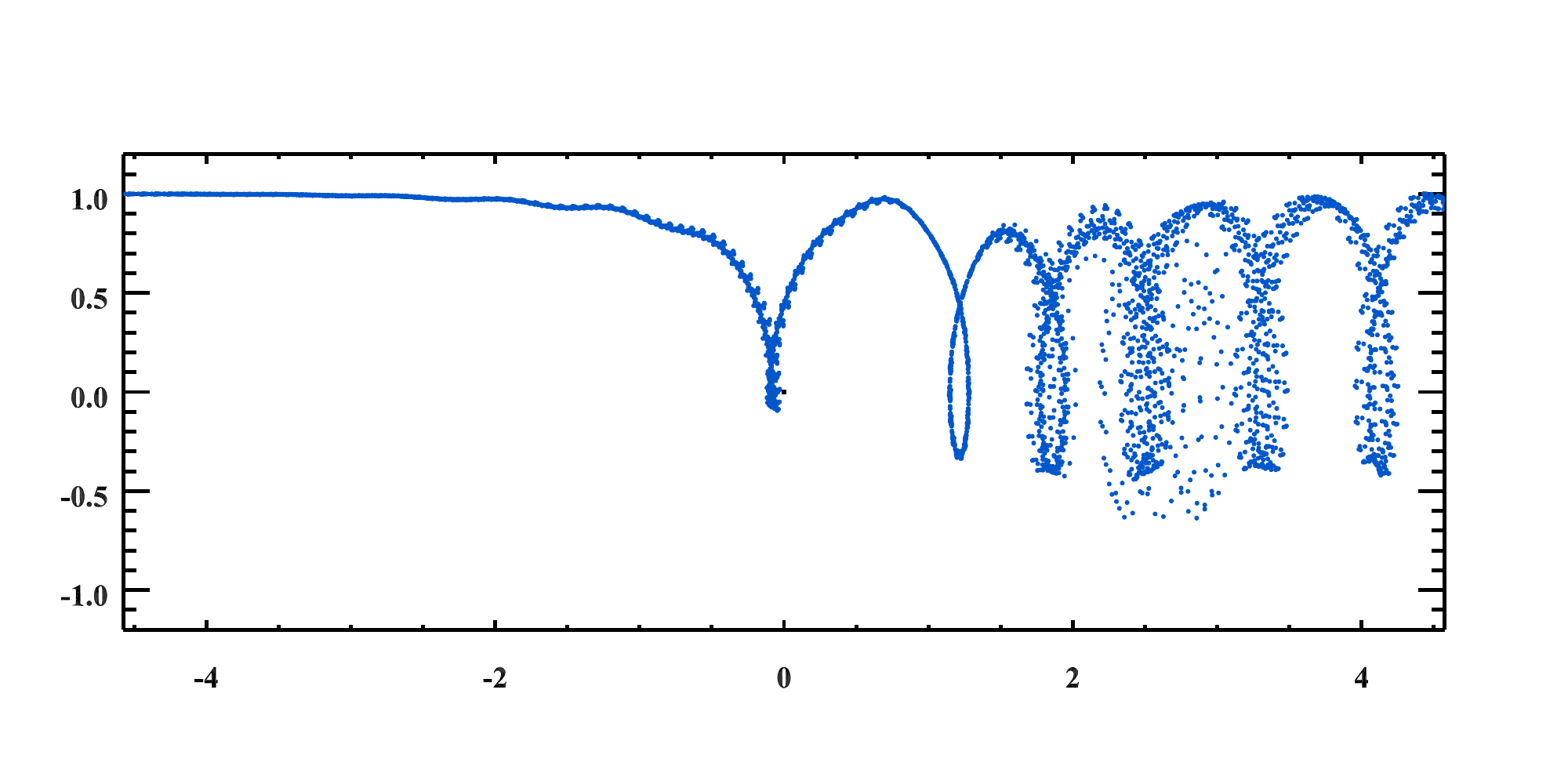}}
	\end{overpic}
	\vskip -11mm
	\begin{overpic}[width=\linewidth]{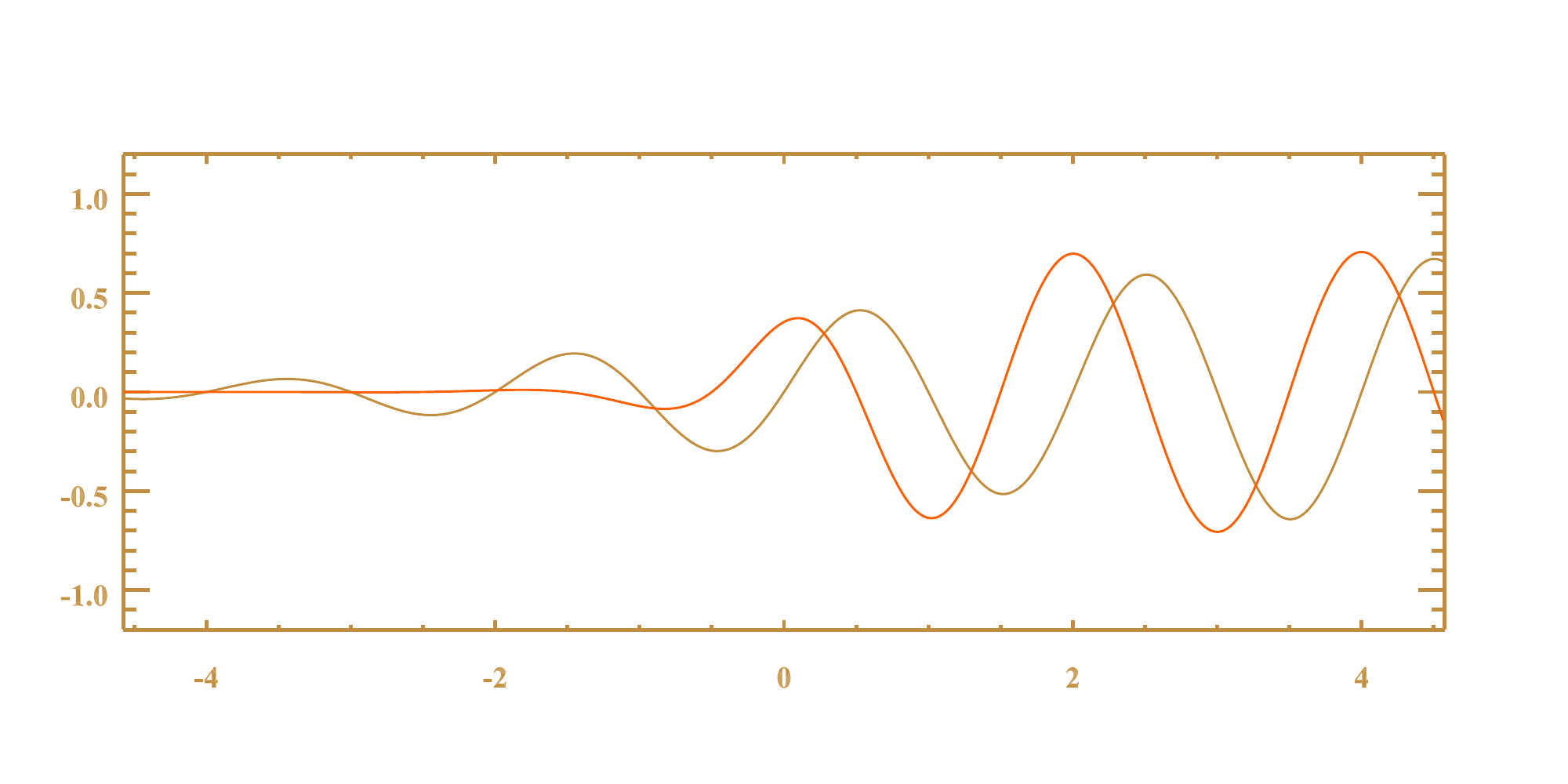}
		\put(0,0){\includegraphics[width=\linewidth]{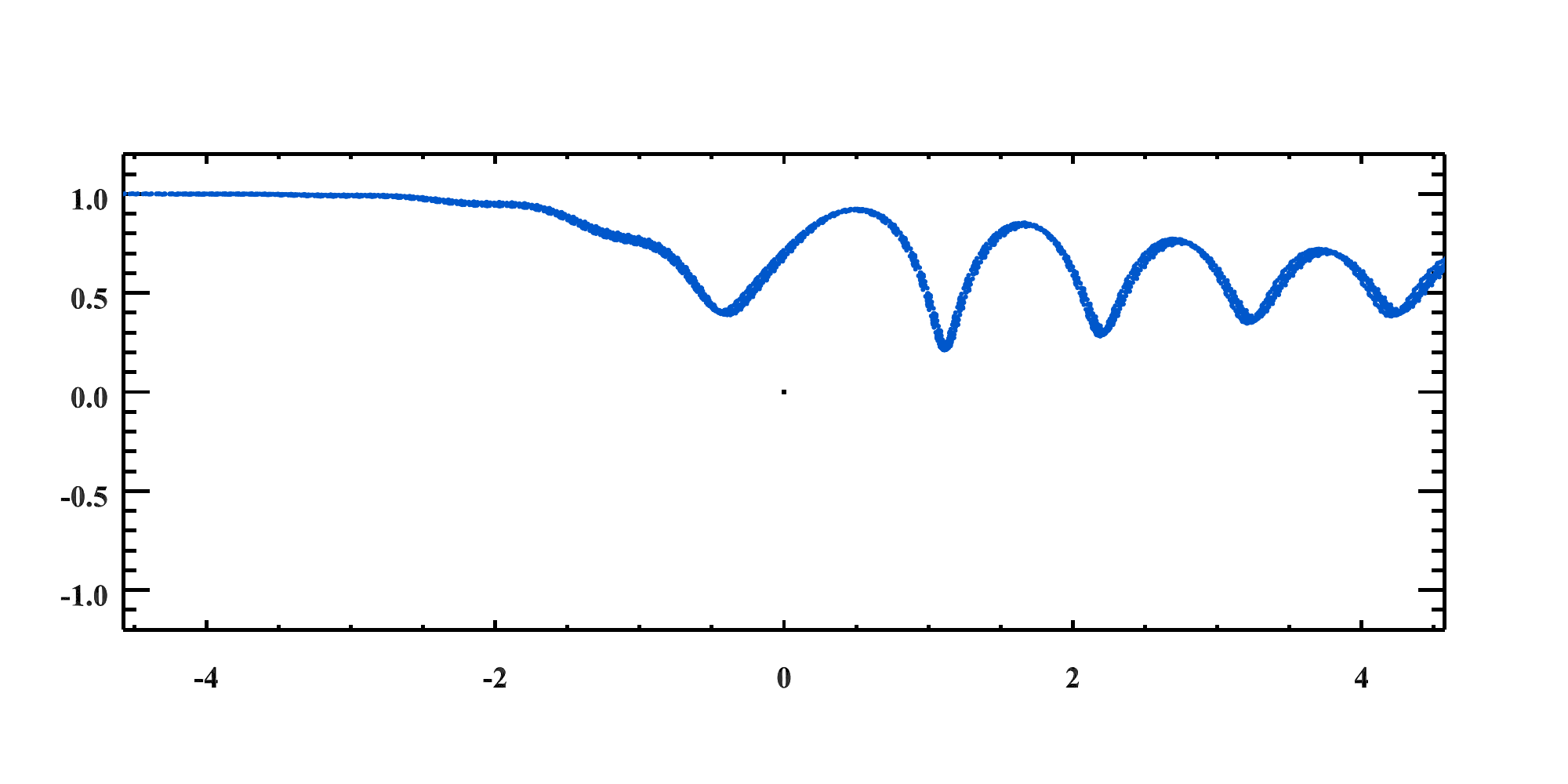}}
	\end{overpic}
	\vskip -3mm
	\caption{Phase space diagrams of instabilities with combined polarizations, interacting with the flow of particles, given for wavelengths near resonance. The transverse components of magnetic field are overplotted as $B_y$ (light brown) and $B_z$ (orange). On the horizontal axes the X-coordinates are given in units of gyro-wavelength ($\lambda_g = \pi r_g$); on the vertical axes, the longitudinal velocities ($\varv_x$) are normalized to $\varv_0$, and the amplitudes of $B_y$ and $B_z$ are normalized to the background field $B_0$. Cases of different instability wavelengths are ordered in succesion from top to bottom, for $\lambda = \{ 0.85,\ 0.90,\ 1.00,\ 1.05,\ 1.50,\ 2.00 \}\ \lambda_g$.}
\end{figure}

Because of such properties, we think that in a steady state, the instability starts to grow at the shock front linearly polarized, and then soon after that, the field lines bend to a circularly polarized shape. The scattering strength and the amount of particles backscattered upstream then depend on the width of the transition region.

\section{Results from PIC simulations and discussion}
\label{discussion}

We observe two populations of particles, quasi-thermal ions which cross downstream and supra-thermal ions which are backscattered to an upstream region. We find that the amount of reflected particles could be significant and similar to case~\citep{NumRefPrtl} where particles are reflected by SDA process, except that non-thermal ions are not present here because the electric field component is omitted (no acceleration) and interaction is provided through resonant scattering processes only. We think that at the beginning, the width of the linear to circular transition could be large ($\sim$ few $\lambda_G$). Its contribution to the initial return current can then be as high as $\sim 40\%$ by number. Later on, such transition region should shrink, thus decreasing the amount of backscattered particles to a few percent -- and if it shrinks completely, backscattering should stop.

Injection by this mechanism could be very important, in a way that it could trigger large return current at the very beginning. This means that short-wavelength \emph{non-resonant} modes, that are driven by this current, could turn on DSA at much earlier stages.

\begin{figure}[!h]
	\centering
	\includegraphics[width=\linewidth]{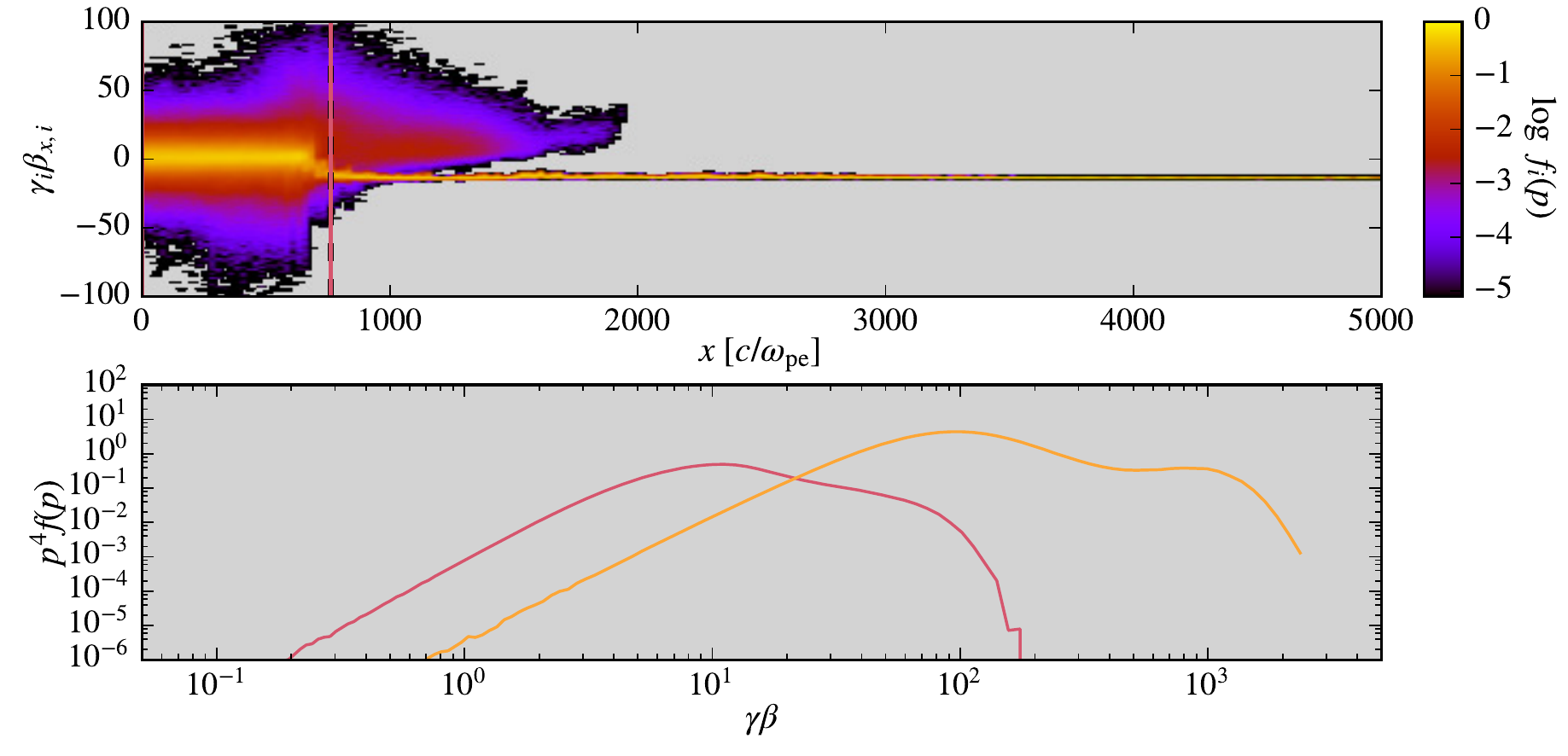}
	\includegraphics[width=\linewidth]{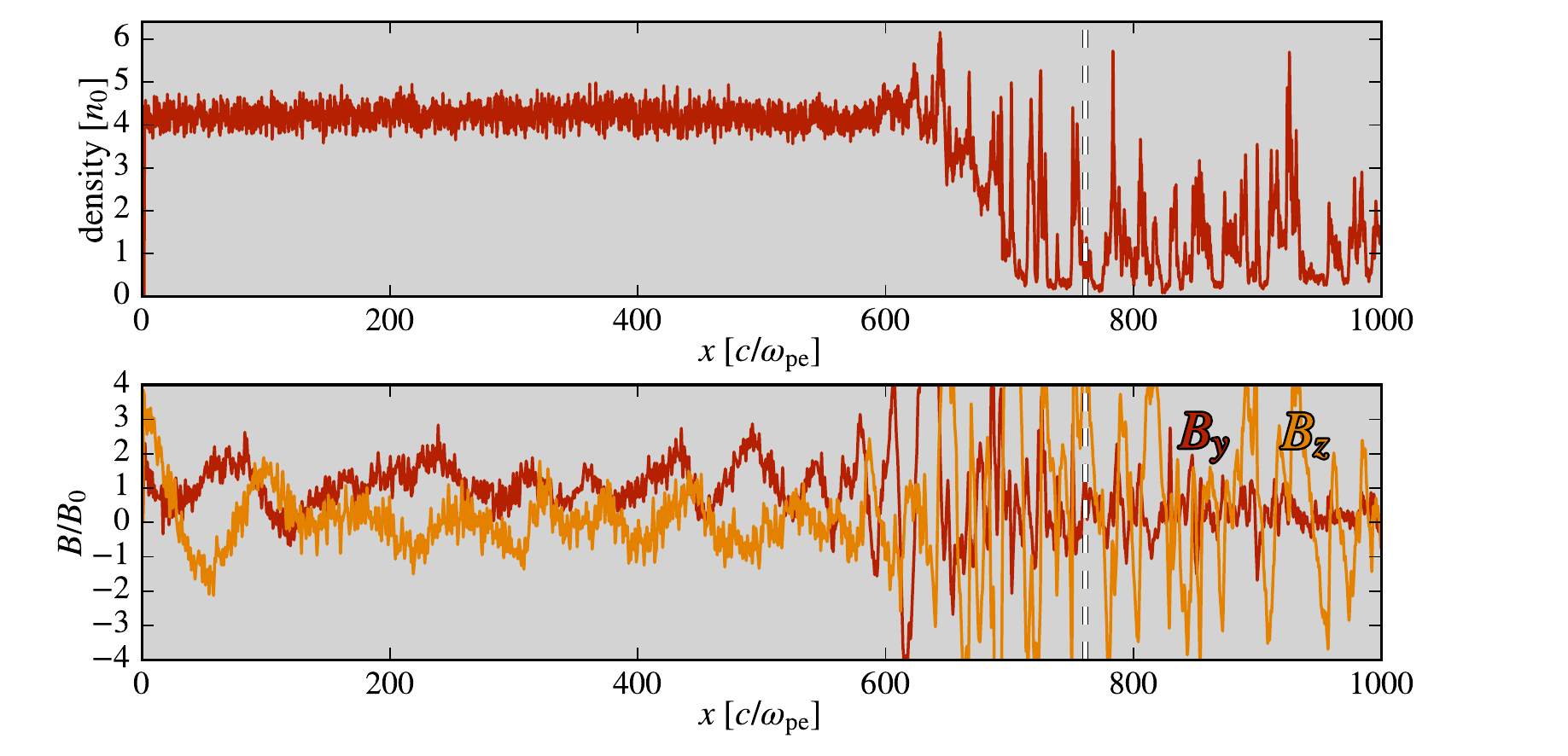}
	\vskip -1mm	
	\caption{Phase space diagram, spectra of particles, density profile and transverse field components $B_y$ and $B_z$ at $t = 2250\ \omega_{pe}^{-1}$.}
\end{figure}

Also, when a \emph{resonant} instability reaches an amplitude $B_0$, scattering starts and the stream of incoming particles is disrupted, and according to (5) it stays frozen in the downstream flow. In this way, \emph{resonant} instabilities might also be responsible for triggering the thermalization process, or could at least contribute to it by isotropizing particle velocities.

Here, we present preliminary results of 2-D simulations with the PIC code TRISTAN-MP \citep{TRISTAN}. The simulation set-up is similar to \citep{SDA}: the bulk Lorentz factor of the electron-ion stream is $\gamma_0 = 15$, the thermal spread is $\Delta\gamma = 10^{-4}$, the magnetization is $\sigma = 0.1$, the magnetic inclination is $\theta = 15^{\circ}$, and the mass ratio is $m_i / m_e = 16$. The computational domain is $\sim 25\ c/ \omega_{pe}$ wide (along $y$) and $\sim 5000\ c/ \omega_{pe}$ long, with a skin depth of 10 cells.

\begin{figure}[!h]
	\centering
	\includegraphics[width=\linewidth]{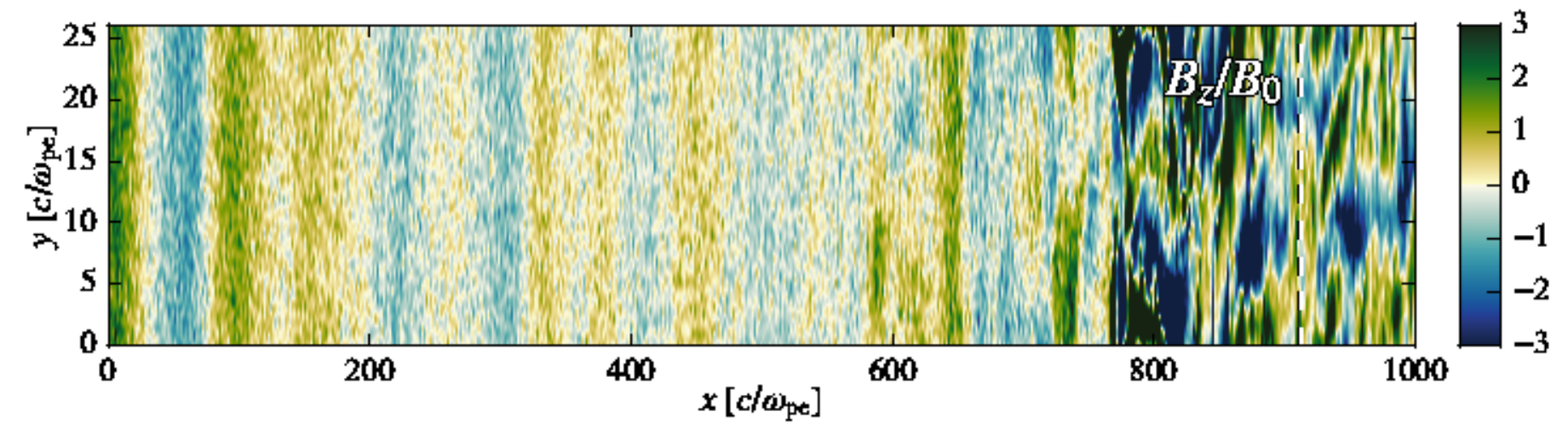}
	\includegraphics[width=\linewidth]{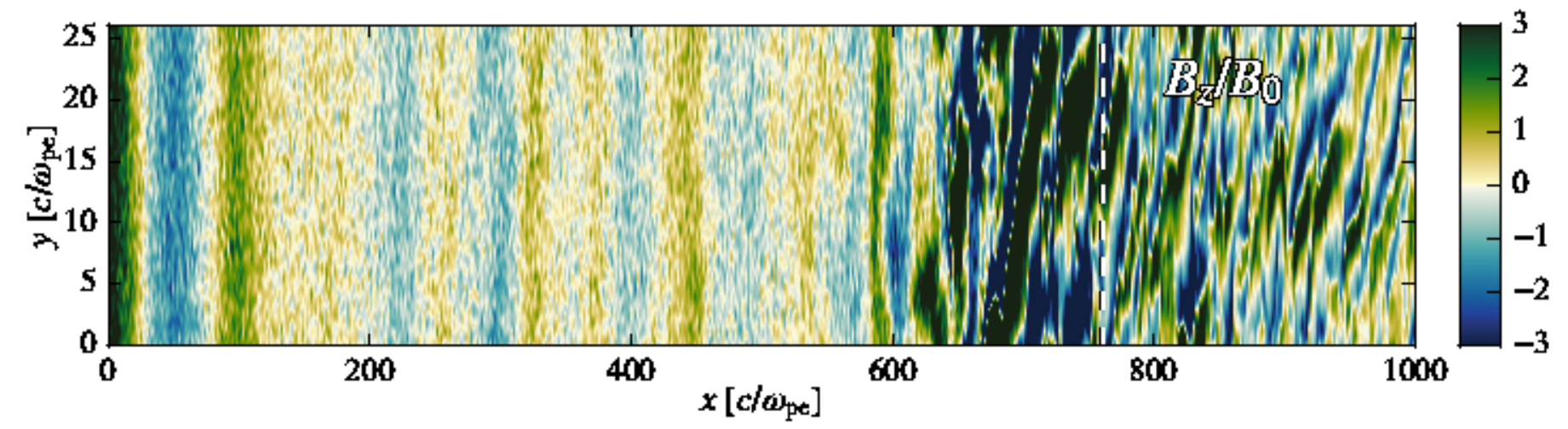}
	\includegraphics[width=\linewidth]{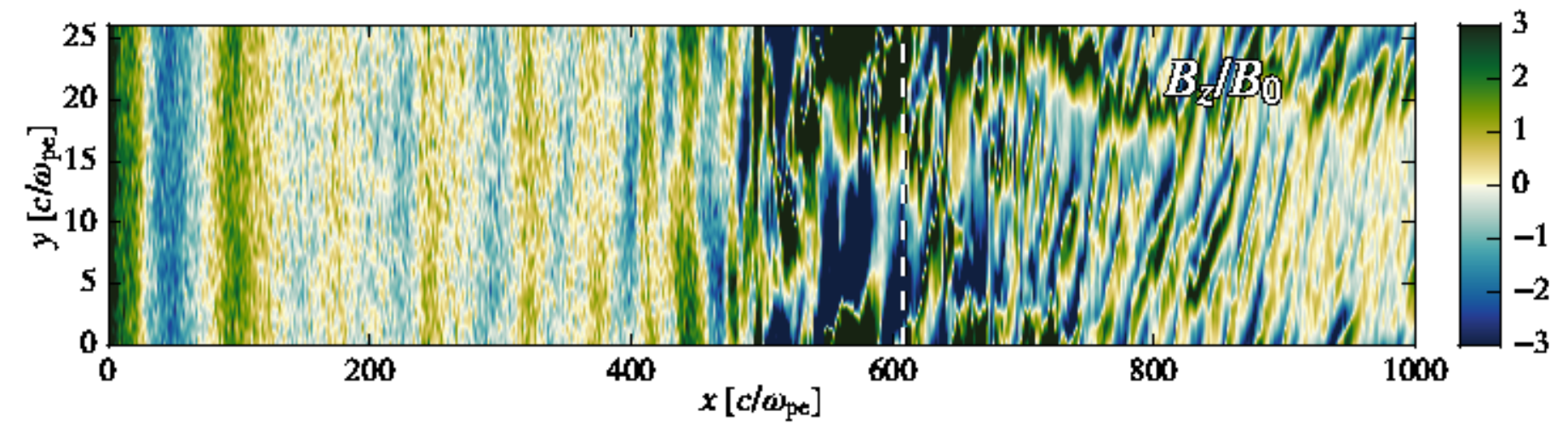}
	\includegraphics[width=\linewidth]{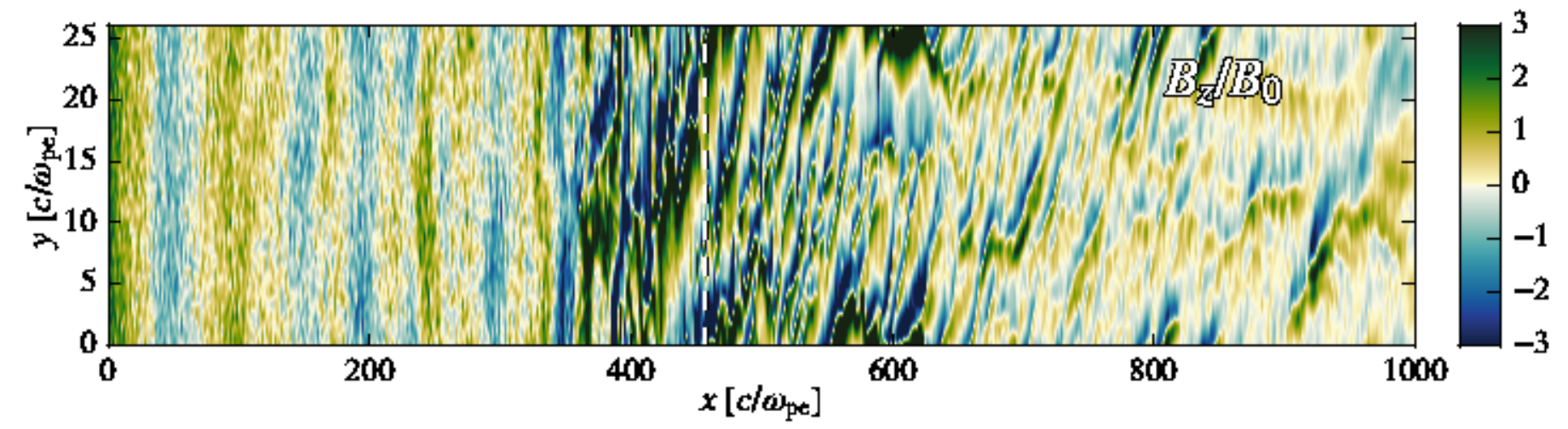}
	\caption{Two-dimensional plot of transverse magnetic field component $B_z$ in units normalized to $B_0$, at times $t = \{ 1350, 1800, 2250, 2700 \} \ \omega_{pe}^{-1}$, from bottom to top respectively.}
\end{figure}

We observe that, circulary polarized \emph{resonant} instabilities grow early on in our simulation (Figure 3). Their wavelength is resonant with the flow, and the amplitude is $\sim B_0$, as in previous calculations. We observe that, when these instabilities grow to $B_0$, a non-thermal peak appears in spectra. As the shock propagates, it imprints these instabilities in a downstream plasma, meaning that they are advected with the downstream flow, and \emph{non-resonant} modes propagate ahead of a shock, as shown in Figure 4.

Finally, we conclude that such \emph{resonant} instabilities could account for many properties of a shock, and contribute in mediating quasi-parallel shocks.

\section*{Acknowledgments}

This research has been supported by the Ministry of Education, Science and Technological Development of the Republic of Serbia
under project No. 176005.




\nocite{*}
\bibliographystyle{elsarticle-num}

\begin{thebibliography}{10}


\bibitem{SDA} Sironi, L., \& Spitkovsky, A., 2011, ApJ, 726, 75
\bibitem{NR} Bell A. R., 2004, MNRAS, 353, 550
\bibitem{RvsNR} Caprioli, D., \& Spitkovsky, A. 2014, ApJ, 794, 46
\bibitem{FDM} Vranjes, J., 2015, Phys. Plasmas, 22, 5, 052102
\bibitem{NumRefPrtl} Caprioli, D., Pop, A-R., Spitkovsky, A., 2015, ApJL, 798, L28
\bibitem{TRISTAN} Spitkovsky, A., 2005, in AIP Conf. Ser. 801, Astrophysical Sources of High Energy Particles and Radiation, 345

\end{thebibliography}



\end{document}